\documentclass{PoS}
\usepackage{epsfig}
\title{
\vspace{-1cm}
\begin{minipage}{\textwidth}
\begin{flushright}
\normalsize DESY-07/171 \\ ITEP-LAT/2007-21 \\ KANAZAWA 07-12 \\
\end{flushright}
\end{minipage}\\[15pt]
Finite temperature Lattice QCD with two flavors of 
improved Wilson fermions}
\ShortTitle{Finite temperature LQCD}

\author{\speaker{V.G.~Bornyakov}~$^{ab}$,
S.M.~Morozov~$^b$,
Y.~Nakamura~$^{cf}$, 
M.I.~Polikarpov~$^b$,
G.~Schierholz~$^{cd}$, and 
T.~Suzuki~$^{ef}$
\\
\llap{$^a$}Institute of High Energy Physics, 142281 Protvino, Russia\\
\llap{$^b$}Institute for Theoretical and Experimental Physics , B.Cheremushkinskaya 25, 117259 Moscow, Russia \\
\llap{$^c$}John von Neumann-Institut f\"ur Computing NIC /
  DESY, 15738  Zeuthen, Germany\\ 
\llap{$^d$}Deutsches Elektronen-Synchrotron DESY,
  22603 Hamburg, Germany\\
\llap{$^e$}Institute for Theoretical Physics, Kanazawa University, Kanazawa 920-1192, Japan \\
\llap{$^f$} RIKEN, Radiation Laboratory, Wako, 351-0158, Japan \\
E-mail: \email{Vitaly.Bornyakov@ihep.ru}, 
\email{Yoshifumi.Nakamura@desy.de}

\vspace{.5cm}

\centering{ \bf \large DIK Collaboration} \\
}

\abstract{We present results of finite temperature QCD from $N_f=2$ flavors of non-perturbatively improved Wilson fermions on lattices with $N_t=8$, 10 and 12. The transition temperature $T_c$ is determined for the first time at lattice spacings as low as $a = 0.08$ fm, albeit at pion masses $1.3 \lesssim r_0 m_\pi$. We furthermore compute the screening masses in various color channels at $T>T_c$.}

\FullConference{The XXV International Symposium on Lattice Field Theory\\

                 July 30 - August 4, 2007\\

                 Regensburg, Germany}

\begin{document}

\section{Introduction}
Despite many efforts to accurately determine the transition temperature $T_c$ in finite temperature QCD, the final answer is still open. The problem is that the lattice data need to be extrapolated to the continuum limit and to the physical quark mass. Indeed, recent work of the RBC-Bielefeld~\cite{RBC-Bielefeld} and the Wuppertal group~\cite{Aoki2006}, using different versions of staggered fermions, report results for $T_c$, which differ by more than 10\%. (See also the plenary talks of Karsch and Fodor in this volume.) To settle this issue, alternative calculations, preferably using different fermionic actions, are demanded. 

In this talk we shall present results for $T_c$ from $N_f=2$ flavors of 
nonperturbatively improved Wilson fermions. (For similar work 
see~\cite{Maezawa2007} and the talk of Namekawa in this volume.) Furthermore, we shall report results for the screening mass, which is controversial as well.

We use the plaquette action. The fermionic action reads
\begin{equation}
S_F = S^{(0)}_F - \frac{\rm i}{2} \kappa\, g\,
c_{sw} a^5
\sum_s \bar{\psi}(s)\sigma_{\mu\nu}F_{\mu\nu}(s)\psi(s)\,,
\label{action}
\end{equation}
where $S^{(0)}_F$ is the standard Wilson action, and $c_{SW}$ is determined nonperturbatively~\cite{Jansen:1998mx}. Here we report preliminary results on $24^3 \times 12$ lattices at $\beta=5.29$, and use our earlier results on $16^3 \times 8$ at $\beta=5.2$, $5.25$ and $24^3 \times 10$ at $\beta=5.2$~\cite{DIK2004,DIK2005}. The transition temperature was determined for lattice spacing $a$ ranging from $0.17 r_0$ to $0.23 r_0$, and the pion mass $r_0 m_\pi$ varying between 1.3 and 2.7. The lattice spacing and pion mass has been determined from the $T=0$ results obtained by the QCDSF-UKQCD collaboration (see e.g. \cite{Gockeler:2006vi}). 

\section{Critical temperature}
\begin{figure}
\centerline{\epsfxsize=14truecm\epsfbox{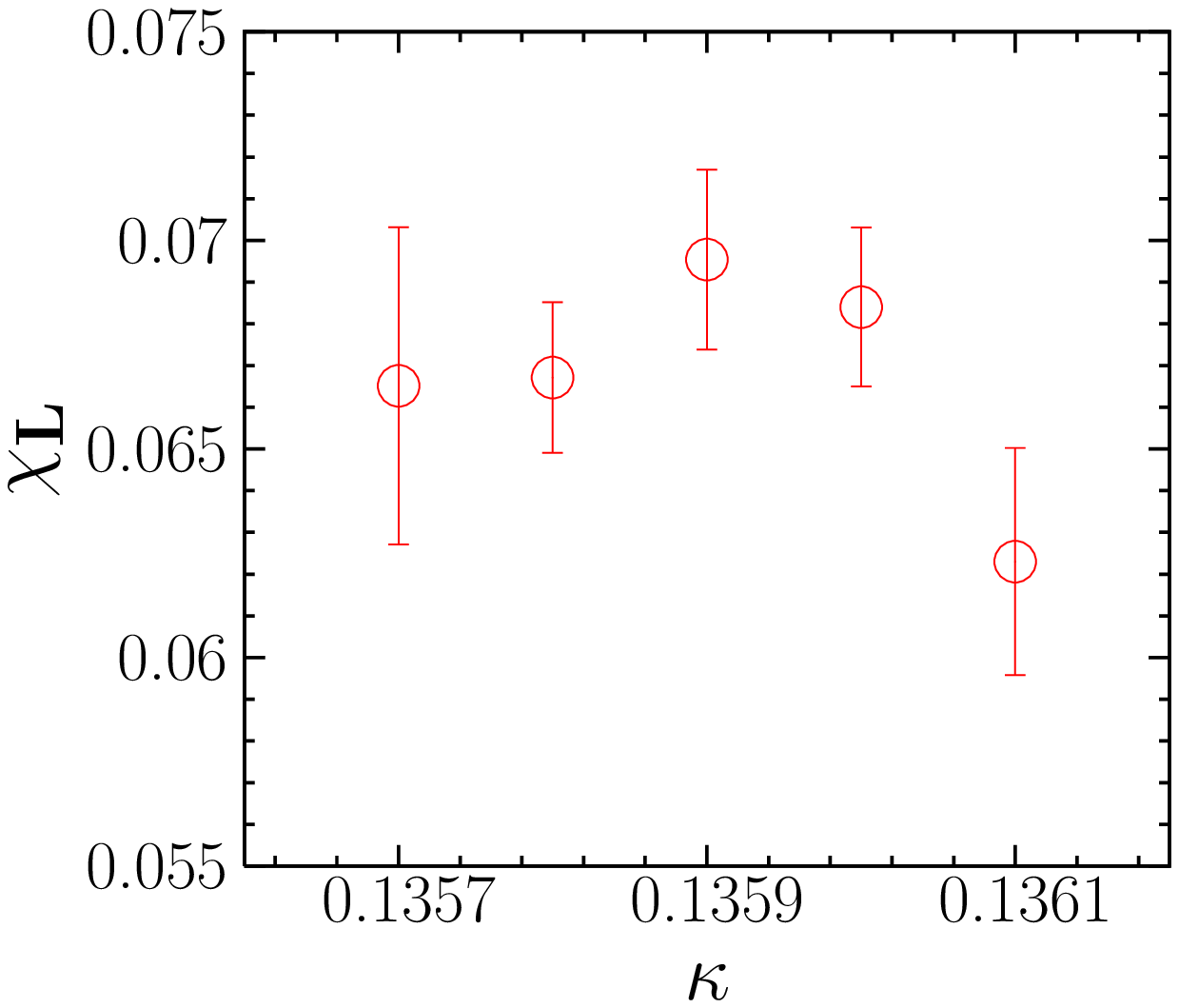}}
\vspace{-9.5cm}
\caption{The Polyakov loop susceptibility from the $24^3 \times 12$ lattice.} 
\label{susceptibility}
\end{figure}

We use the Polyakov loop susceptibility $\chi_L$ to compute $T_c$. In Fig.~\ref{susceptibility} we show our results on the $24^3 \times 12$ lattice.
The critical value of $\kappa$, $\kappa_t$, is identified as a point, where $\chi_L$ reaches its maximum. 
Applying a Gaussian fit in the vicinity of the maximum we find
$\kappa_t=0.13589(6)$. Respective value of $T_c r_0$ is $0.487(6)$

To determine $T_c$ in the continuum limit and at the physical pion mass we fit the mass and cut-off dependence by
\begin{equation}
r_0 T_{c}(r_0 m_\pi,1/N_t) = r_0 T_{c}(0,0)
                             + c_N \cdot \frac{1}{N_t^2}
                             + c_m \cdot (r_0 m_\pi)^d
\label{fitting1}
\end{equation}
with $d=1.08$, assuming that the transition is of second order in the chiral limit and is in the universality class of the $3d~O(4)$ spin model. Note that a first order transition at the physical quark masses is not fully excluded~\cite{D'Elia2005}. In that case $d=2$. We treat the difference between the two fits with $d=1.08$ and $d=2$ as a systematic error. We also considered an extrapolation of the form
\begin{equation}
r_0 T_{c}(r_0 m_\pi,a/r_0) = r_0 T_{c}(0,0) 
                             + c_a \cdot \left( \frac{a}{r_0}\right)^2 
                             + c_m \cdot (r_0 m_\pi)^d \,.
\label{fitting2}
\end{equation}
Again, the difference between (\ref{fitting1}) and (\ref{fitting2}) was taken as a systematic error.

\begin{figure}
\label{crit_temp}
\centerline{\epsfxsize=14truecm\epsfbox{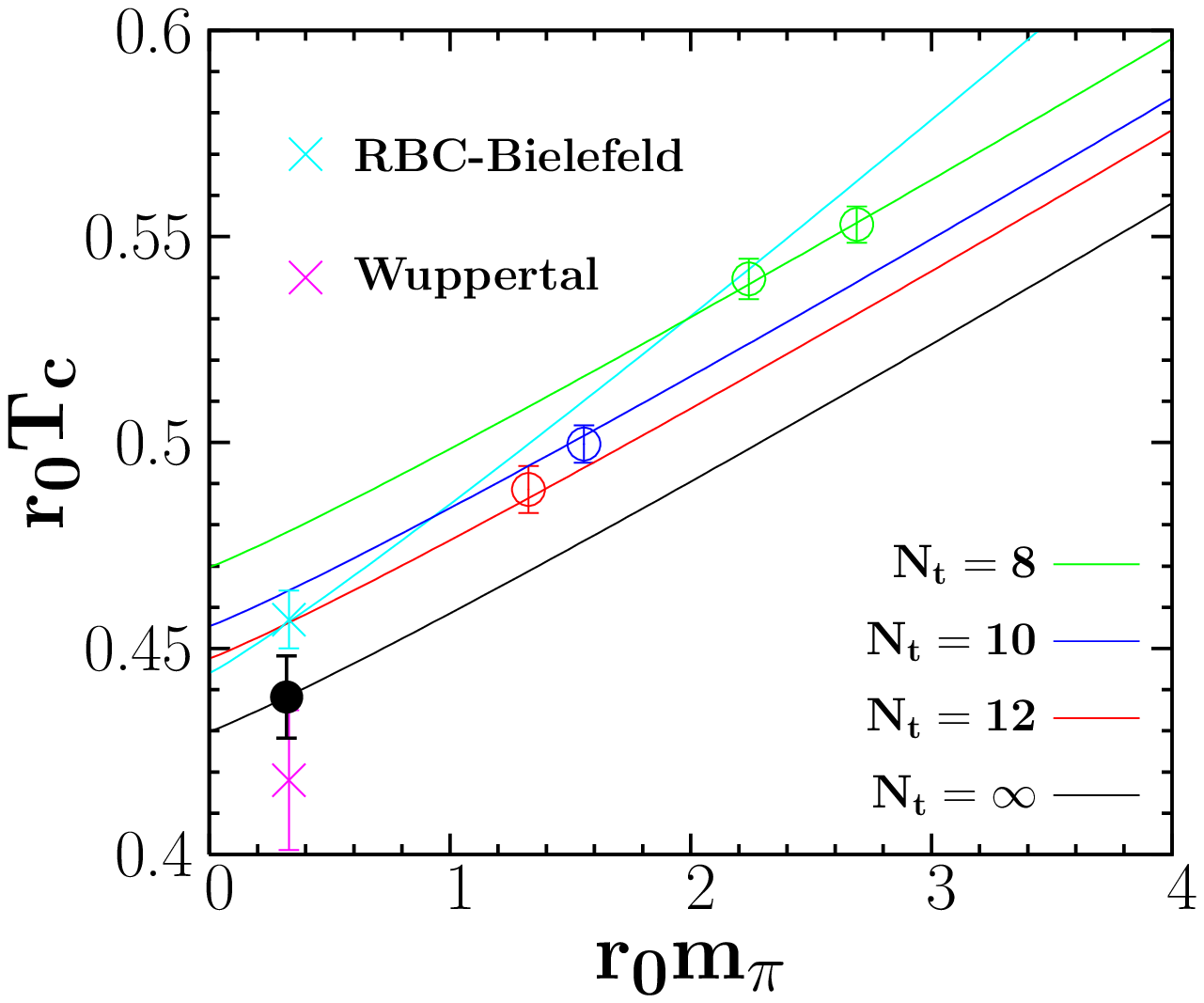}}
\vspace{-9.5cm}
\caption{The critical temperature. The solid lines show results of the fit function 
(2.1).
The blue line shows the fit of~\cite{RBC-Bielefeld}.}
\end{figure}

Our results are presented in Fig.~\ref{crit_temp}, where we show the fit
function (\ref{fitting1}) for $N_t=8, 10, 12$ and $N_t=\infty$. It turns out
that the coefficients $c_N$ (respectively $c_a$) and $c_m$ are strongly
correlated. The error on (e.g.) $c_N$ is approximately $50\%$, so that the
scaling violations might be significantly smaller than displayed in
the figure. At present only the result in the physical limit can be trusted
entirely. 
In the continuum limit and at the physical pion mass we obtain 
\begin{equation}
 r_0 T_{c}(r_0 m_\pi^{ph},0) = 0.438(6)(^{+13}_{-7}) \,,
\label{crit_temp_eq}
\end{equation}
where the first error is statistical and the second one systematic.

For comparison we also show the fit function of~\cite{RBC-Bielefeld}. It would
be consistent with our results (within the error bars) if $c_N=0$. The result
of our fit (\ref{fitting1}) is $2\sigma$ away from that value. 

As we said already, we used the Polyakov loop susceptibility to compute
$T_c$. Another observable, the chiral susceptibility, determined
from the low lying eigenmodes of the valence overlap Dirac operator, was
computed on our $24^3 \times 10$ lattice in the parallel talk of V. Weinberg
in this volume. The preliminary conclusion was that the respective transition
temperature is shifted to a lower value as compared to the Polyakov loop
susceptibility. This observation is in agreement with the findings of Ref.\cite{Aoki2006}.

\section{Screening masses at $T>T_c$}
The study of free energies of static quarks in the high temperature phase is important for understanding of the fate of charmonia above $T_c$, as well as for checking the validity of high temperature perturbation theory.
The free energy in the different color channels is given by~\cite{McLerran}:
\begin{eqnarray}
\label{definitions}
e^{-F_1(R,T)/T} &=& \frac{1}{3} \langle ~\mbox{Tr} L^\dagger(x) L(y) ~\rangle \\ 
\nonumber
e^{-F_8(R,T)/T} &=& \frac{1}{8} \langle  ~\mbox{Tr} L^\dagger(x) ~\mbox{Tr} L(y)~\rangle - \frac{1}{24} \langle ~\mbox{Tr} L^\dagger(x) L(y) ~\rangle\\ \nonumber
e^{-F_6(R,T)/T} &=& \frac{1}{12} \langle  ~\mbox{Tr} L(x) ~\mbox{Tr} L(y)~\rangle + \frac{1}{12} \langle ~\mbox{Tr} L(x) L(y) ~\rangle\\ \nonumber
e^{-F_3^*(R,T)/T} &=& \frac{1}{6} \langle  ~\mbox{Tr} L(x) ~\mbox{Tr} L(y)~\rangle - \frac{1}{6} \langle ~\mbox{Tr} L(x) L(y) ~\rangle
\end{eqnarray}
The above definitions are not explicitly gauge invariant. It was argued in 
Ref.~\cite{Philipsen2002} that a gauge invariant result can be obtained after
proper gauge fixing. Although a rigorous proof is still lacking and
both  definitions (\ref{definitions}) and gauge invariance were questioned in
Ref.~\cite{Philipsen2004} and  Ref.~\cite{BBM2003}, respectively, we, as most of other authors, choose here the Coulomb gauge to study potentials determined by eqs.~(\ref{definitions}). An iterative gauge fixing algorithm with one gauge copy has been applied. It is known that in the Coulomb gauge
some gauge noninvariant quantities, e.g. field propagators, suffer from the
Gribov problem. To check for the effect of Gribov copies we looked at three
copies on a subset of our configurations but found no effect. Still, further,
more extensive, checks of these effects  are necessary. In particular, a more effective  simulated annealing algorithm should be applied.

To improve the signal to noise ratio we used a hypercubic blocking
procedure~\cite{Hasenfratz2001}. This reduced the statistical errors by about
a factor of three.

We applied the usual fitting function to describe the static potentials as functions of $R$ at large $RT$:
\begin{equation}
\nonumber
 V_i(R,T) \equiv F_i(R,T) - F_i(\infty,T) = -C_i \frac{\alpha_i(T)}{R}
 e^{-m_{D \, i}(T)R}
\label{potential}
\end{equation}
($i=1, 3, 6, 8$), where  $\alpha_i(T)$ and $m_{D\,i}(T)$ are fit parameters, and
$C_i$ is the Casimir factor. 

Previous studies in $N_f=2$  lattice QCD were performed with
staggered fermions \cite{Kaczmarek2005} as well as  with improved Wilson
fermions~\cite{WHOT-QCD2007}, in both cases on lattices with $N_t=4$. 
In Ref.~\cite{Kaczmarek2005} results were presented for  $m_\pi/m_\rho = 0.7$
and temperatures up to $T/T_c=4$. These authors fitted the screening
mass $m_D$, extracted from the singlet static potential by means of
(\ref{potential}), in terms of the two-loop perturbative expression times a
constant $A$ to allow for nonperturbative corrections:  
\begin{equation}
\nonumber
  \frac{m_D}{T} = A\left( 1+\frac{N_f}{6} \right) ^{1/2} g_{two-loop}(T). 
\end{equation}
They found $A \approx 1.4 $, to be compared with $A=1$ in perturbation theory.

In Ref.~\cite{WHOT-QCD2007} approximately the same values of $m_\pi/m_\rho$   ($m_\pi/m_\rho = 0.65, 0.80$) and same temperature range were explored.
The authors confirmed Casimir  scaling for $V_M(R,T)$, i.e. independence of 
$\alpha_i(T)$ and $m_{D\,i}(T)$ on $i$. Furthermore, they found the phenomenological relation:
\begin{equation}
\nonumber
\frac{m_D}{T} =  \left( 1+\frac{N_f}{6} \right) ^{1/2} \sqrt{4\pi\alpha(T)}
\label{scaling}
\end{equation}
Comparison with results of Ref.~\cite{Kaczmarek2005} revealed agreement for
$\alpha(T)$, but  20\% deviation for $m_D(T)$. So far the disagreement of $m_D(T)$ is unclear. One possible reason is that in both cases lattices with large lattice spacing were used.
Thus our result obtained on lattice with small lattice spacing can shed some light on
this problem.

\begin{figure}
\centerline{\epsfxsize=14truecm\epsfbox{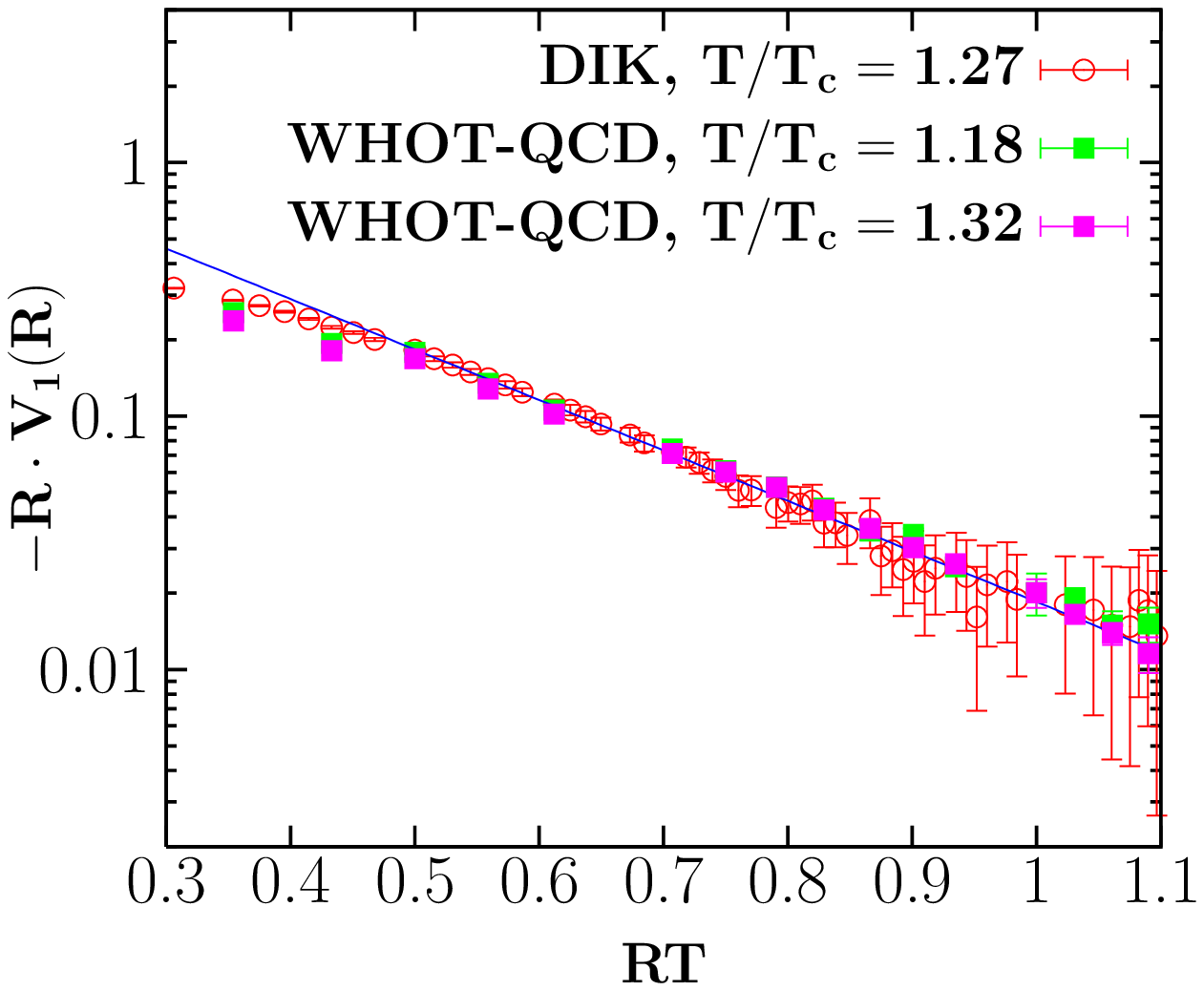}}
\vspace{-10cm}
\caption{Comparison with \cite{WHOT-QCD2007} for the singlet potential.} 
\label{singlet}
\end{figure}

Our calculations of the static potentials are performed on $16^3 \times 8$
lattice at $T/T_c=1.27$, the highest available temperature. In
Fig.~\ref{singlet} we compare our result for  $V_1(R,T)$  with that of
Ref.~\cite{WHOT-QCD2007}. We find good agreement for $RT > 0.5$. The
disagreement at smaller distances might be partially due to breaking of
rotational invariance and HYP smearing of the data of
Ref.~\cite{WHOT-QCD2007}.  
Similar agreement was found for other color channels. 
Thus we confirm the observations made in \cite{WHOT-QCD2007}, i.e Casimir scaling and phenomenological scaling (\ref{scaling}).

\section{Conclusions}
We  computed $T_c$  for $N_f=2$ with improved Wilson fermions on lattices with
$N_t = 8, 10, 12$ and performed a continuum and chiral extrapolation. The
physical value of $r_0 T_c$ was found to be in agreement with both the
RBC-Bielefeld and Wuppertal results within the error bars.  

The screening masses at $T/T_c \approx 1.3$ are found in full agreement
with WHOT-QCD results \cite{WHOT-QCD2007}, confirming disagreement with the
results of staggered fermions~\cite{Kaczmarek2005}, as well as Casimir scaling of $V_i(R,T)$ and phenomenological scaling of $m_D(T)$.

\acknowledgments
This work is supported by the Large Scale Simulation Program No.07-14-B of
High Energy Accelerator Research Organization (KEK).
The numerical simulations were done using RSCC at RIKEN and BlueGene/L at
KEK.
V.G.B., S.M.M. and M.I.P. are supported by grants RFBR  05-02-16306, 07-02-00237-a, RFBR-DFG 06-02-04010, DFG-RFBR 436 RUS 113/739/0-1 and by the EU Integrated Infrastructure Initiative Hadron Physics (I3HP) under contract RII3-CT-2004-506078.
S.M.M. is also supported by an INTAS~YS fellowship 05-109-4821.

\end{document}